\documentclass[%
reprint,
amsmath,amssymb,
aps,
floatfix,
]{revtex4-1}

\usepackage{graphicx}
\usepackage{dcolumn}
\usepackage{amsmath}
\usepackage{amssymb}
\usepackage{bm}
\usepackage[bbgreekl]{mathbbol}
\usepackage{color} 
\usepackage{braket}
\usepackage{gensymb}
\usepackage[dvipsnames]{xcolor}
\usepackage{lpic}

\usepackage[percent]{overpic}
\usepackage{stackengine}
\usepackage{physics}
\usepackage{siunitx}
\usepackage{textcomp}
\usepackage{slashbox}
\usepackage{lipsum}
 \usepackage{hyperref}
 \hypersetup{
    colorlinks=true,
    linkcolor=blue,
    filecolor=magenta,
    citecolor=blue,
    urlcolor=cyan,
}

\usepackage[mathscr]{eucal}

\usepackage{amsfonts}
\usepackage[colorinlistoftodos]{todonotes}
\usepackage{mathtools}
\usepackage{empheq}
\usepackage{esvect}
\usepackage{rotating}

\DeclareSymbolFontAlphabet{\mathbbm}{bbold}
\DeclareSymbolFontAlphabet{\mathbb}{AMSb}

\begin{document}

\preprint{APS/123-QED}

\title{Quantum noise correlations of an optical parametric oscillator based on a non-degenerate four wave mixing process in hot alkali atoms}
\author{A. Monta\~na Guerrero$^1$}
\author{P. Nussenzveig $^1$}
\author{M. Martinelli $^1$}
\author{A. M. Marino $^2$}
\author{H. M. Florez $^1$}
 \email{hans@if.usp.br}
\affiliation{$^1$ Instituto de F\'{\i}sica, Universidade de S\~ao Paulo, 05315-970 S\~ao Paulo, SP-Brazil}
\affiliation{$^2$ Center for Quantum Research and Technology and Homer L. Dodge Department of Physics and Astronomy, The University of Oklahoma, Norman, Oklahoma 73019, USA}

\date{\today}

\begin{abstract}
We present the first measurement of two-mode squeezing between the twin beams produced by a doubly resonant optical parameter oscillator (OPO) in above threshold operation, based on parametric amplification by non degenerate four wave mixing with rubidium $^{85}$Rb. 
We demonstrate a maximum intensity difference squeezing of -2.7 dB (-3,5 dB corrected for losses)  with a pump power of 285 mW and an output power of 12 mW for each beam, operating close to the D1 line of Rb atoms. 
The possibility to use open cavities combined with the high gain media can provide a strong level of noise compression, and the access to new operation regimes that could not be explored by crystal based OPOs.
The spectral bandwidth of the squeezed light is broadened by the cavity dynamics, and the squeezing level is robust for strong pump powers. Stable operation was obtained up to four times above the threshold. Moreover, its operation close to the atomic resonances of alkali atoms allows a natural integration into quantum networks including structures such as quantum memories.

\end{abstract}

\pacs{Valid PACS appear here}
\maketitle


The generation of quantum correlated fields
is a fundamental resource for developing a quantum network for quantum communication and quantum computation \cite{Nielsen11}. In the context of continuous variables, optical parametric oscillators (OPO) have become the keystone to engineer correlated and entangled light beams on solid state platforms, both below \cite{Ou92} and above \cite{Villar05} its oscillation threshold. Further interest also comes from the development of quantum correlated light sources in optical chips \cite{Avik15}. OPOs have also been used to create large ensembles of multimode entangled fields~\cite{Pysher11,Pinel12,Shota13,Moran14}. 
Moreover, it would be interesting to have these sources operating at wavelengths that are compatible with alkali atoms, which are good candidates for quantum memories or registers~\cite{Lvovsky09}.

On the other hand, 
it was shown that alkali atoms in vapour cells
can generate quantum correlated beams at the atomic wavelength with high quantum correlations~\cite{PDLett07} by parametric amplification using four wave mixing (4WM) based on the third order non linearity $\chi^{(3)}$.
The operation requires a relatively strong pump power beam and an extra seed produced by an acousto-optical modulator.
One of the main difference of this parametric process is the high gain (from 2 to 20 fold) to produce the phase insensitive amplification of the Stokes and anti-Stokes fields \cite{Glorieux10}, when compared to typical gain obtained from parametric amplification by $\chi^{(2)}$ or $\chi^{(3)}$ processes in crystals.

The combination of the high gain amplifier with a cavity could lead to interesting dynamics, provided by the low threshold power that could be obtained, or the study of extreme regimes of open cavities or strong pump operation. 
However, the best gain is reached with a non colinear coupling between the pump and Stokes (anti-Stokes) probe field, in order to satisfy the optimal phase matching condition.
Turnbull et al. studied a range of angles for a proper phase matching condition to obtain high gain on a typical 4WM process~\cite{Turnbull13}. The quantum correlations in a 4WM is
drastically reduced for smaller angles between the probe and conjugate beams, and it was shown that below 2 mrads the intensity difference  squeezing is lost~\cite{Rong18}. 

OPOs using 4WM in atoms have  been recently reported ~\cite{Okuma09,Xudong10,Sheng12}, employing a vapour cell with natural abundance within a cavity to run an OPO above threshold with twin beams separated by 6.1 GHz (for $^{85}$Rb) and 13.6 GHz (for $^{87}$Rb). Some correlations of quantum nature where observed  in the photon counting regime~\cite{Wu09}. For the continuous variables of the field, twin beams were observed for a seeded single resonant OPO with an open cavity for the conjugate mode \cite{SROPO}. Nevertheless, there has not been demonstrations of quantum correlations in self-oscillating cavities. In fact, refs.~\cite{Okuma09,Sheng12}, have reported measurements of intensity noise correlations, but they were not strong enough to reach the quantum limit.

Our purpose is to demonstrate the
generation of quantum  correlated beams from a doubly resonant OPO above the oscillation threshold, using a  $\chi^{(3)}$ order interaction by a non-degenerate 4WM process  employing a hot vapour cell of alkali atoms within a cavity.

\begin{figure}[h!]
\centering
\begin{overpic}[width=86mm]{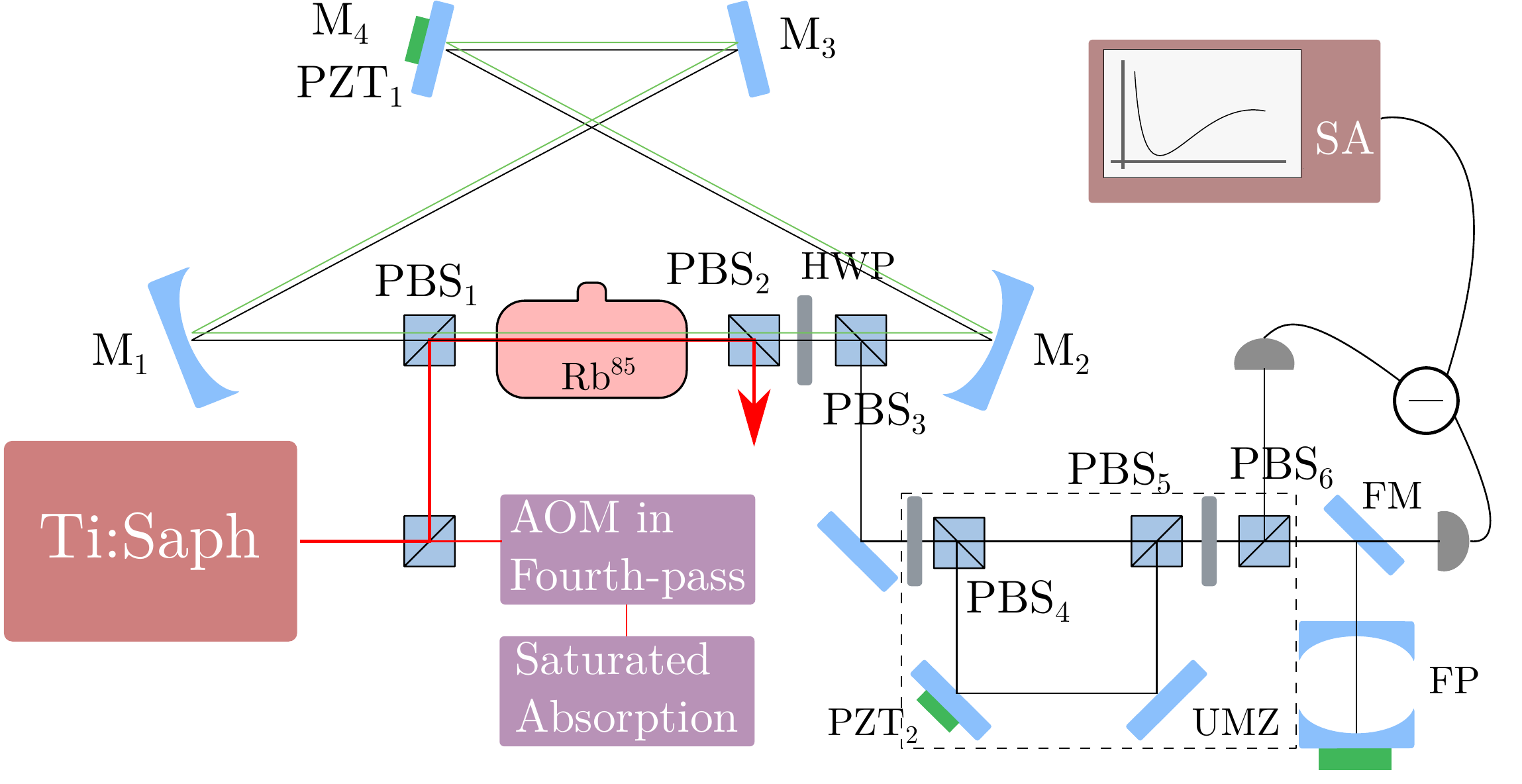}
\put(0,45){(a)}
\end{overpic}
\begin{overpic}[width=86mm]{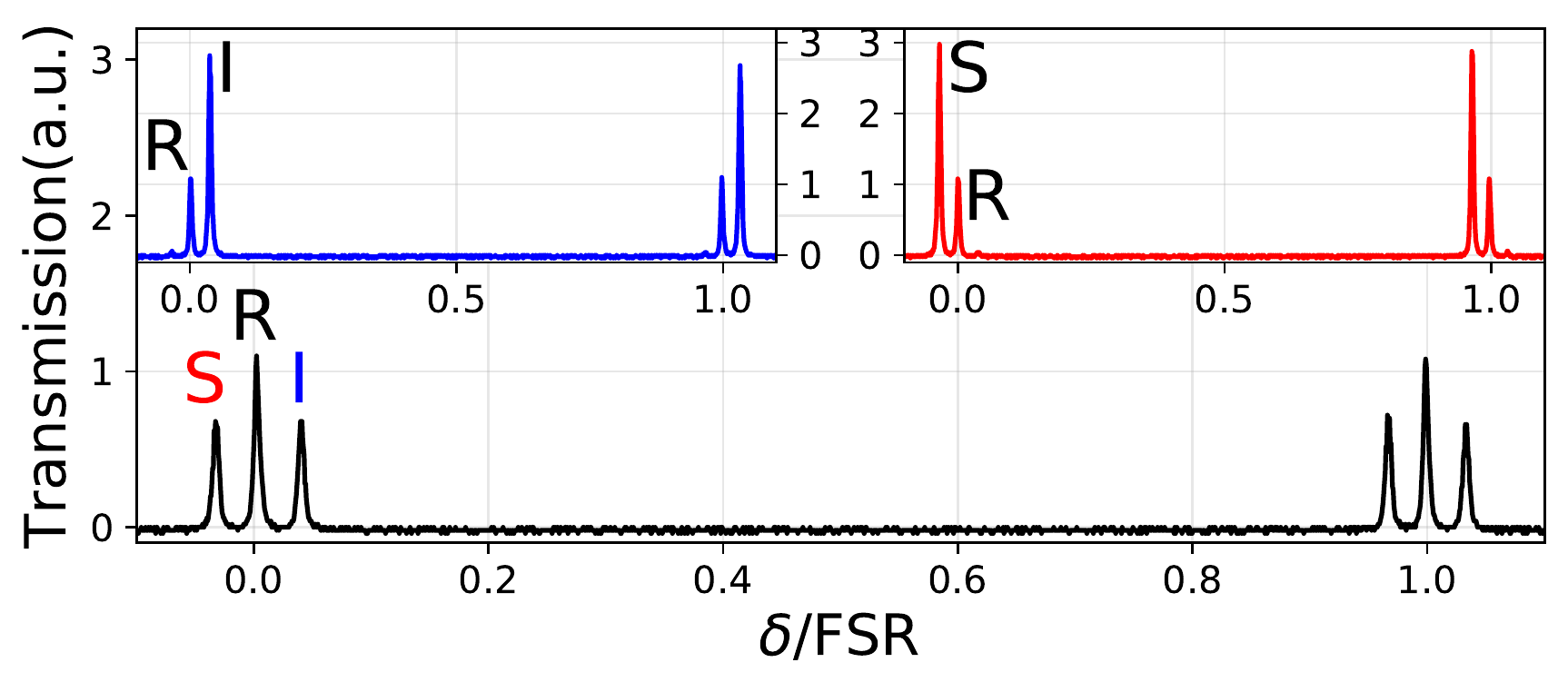}
\put(0,45){(b)}
\end{overpic}
\caption{(a) Sketch of the experimental setup. PBS indicates polarising beamsplitter cube, HWP half-wave plate, UMZ unbalanced Mach-Zehnder interferometer,M mirror, FP confocal Fabry-Perot,  FP flip-mirror and SA spectrum analyser. (b) Interferometer output analysed by a FP cavity with R,S and I  as reference, signal and idler, respectively. Inset: spectral separation of the signal and idler beams. }
\label{fig:Setup}
\end{figure}

The experimental setup is shown in Fig.~\ref{fig:Setup}. We employ a bow-tie cavity  with 4 mirrors with 99.5\% reflectivity, where two of them (M$_1$, M$_2$) have a curvature radius of 50 cm and the other two are flat mirrors (M$_3$, M$_4$). Cavity length is finely controlled by the displacement of mirror M$_4$, mounted on a piezoelectric actuator
(PZT$_1$). 
The effective cavity size of 69.70 (5) cm implies a free spectral range (FSR) of 404 MHz, satisfying the simultaneous resonance for the generated fields, whose frequency separation of 6.070 GHz corresponds to twice the hyperfine splitting of the ground state of the $^{85}$Rb atoms.
The incidence angle of 11$^\circ$ to the normal of the cavity mirrors leads to a negligible astigmatism of the cavity Gaussian mode ($< 3\%$ on the vapour cell). The value of the two waist radius are 316 (3) $\micro m$ and 193 (5) $\micro m$ for the longer and shorter arms, respectively.
In the waist of the longer arm we have a 3 cm long vapour cell with $^{85}$Rb isotope. The vapour cell has anti-reflection coated windows and is kept at 90$^\circ$C for high optical density.
 The pump beam is colinear with the cavity mode and is injected by polarising beam splitter PBS$_1$ and removed by  PBS$_2$.
 After PBS$_2$ we use  a half wave-plate (HWP) and a third PBS (PBS$_3$) in order to control the output coupling of the cavity. By changing the orientation of the waveplate, the cavity finesse ranges from 5 to 30 for a field far from the atomic resonance.

The  pump beam is generated by a Titanium Sapphire laser tuned at 795nm, resonant to the D1 line of Rb and with maximum power of $~$800 mW.
The laser beam is locked to the blue of the $5^2S_{1/2}F=2\rightarrow 5^2P_{1/2}F=3$ transition, with a detuning $\Delta = $1.00, 0.82 or 0.64 GHz. The locking is performed by shifting the frequency of a sample of the laser with an acousto-optic modulator (AOM), driven by a frequency of 250 MHz, on a four pass scheme. The frequency of the shifted field is analysed by saturated absorption spectroscopy, that provides the error signal for the laser frequency stabilization.

Since the  the twin beams are generated with the same polarisation, they need to be  spectrally discriminated.
We used an unbalanced Mach-Zehnder (UMZ) with a difference of optical path of $\sim$25 mm between both arms, as in ref. \cite{Huntington05}. The beams are split by the half waveplate and PBS$_4$, then they are overlaped on  PBS$_5$. By using a HWP and PBS$_6$, we obtain two outputs, with a fringe visibility of 99\%. 
Phase shift of the longer arm is finally adjusted by displacing one of the mirrors with PZT$_{2}$.
This visibility gives us ~98\% separation efficiency for the two modes from the OPO. This can be  verified by using a confocal Fabry-Perot cavity (FP) of 1.5 GHz of FSR (see inset Fig. \ref{fig:Setup}.(b)). Taking an injected reference field (R) from the pump beam, we can adjust the length of the long path for a constructive interference of the signal (S) or the idler field (I), while verifying that their frequencies correspond to those of the hyperfine splitting.

Once the beams are separated, they are sent to two  p-i-n photodiodes (FND-100, EG\&G). The average value of the photocurrents are measured, and the high-frequency component is strongly amplified by a transimpedance circuit. The high frequency signals
are added or subtracted 
and measured by a spectrum analyser (SA), with a resolution bandwidth RBW of 100 kHz and a video bandwidth VBW of 1 kHz. We have an overall detection efficiency of 83 \% for the whole system, accounting for optical losses and photodetector quantum efficiency.

\begin{figure}[h!]
\centering
\begin{overpic}[width=86mm]{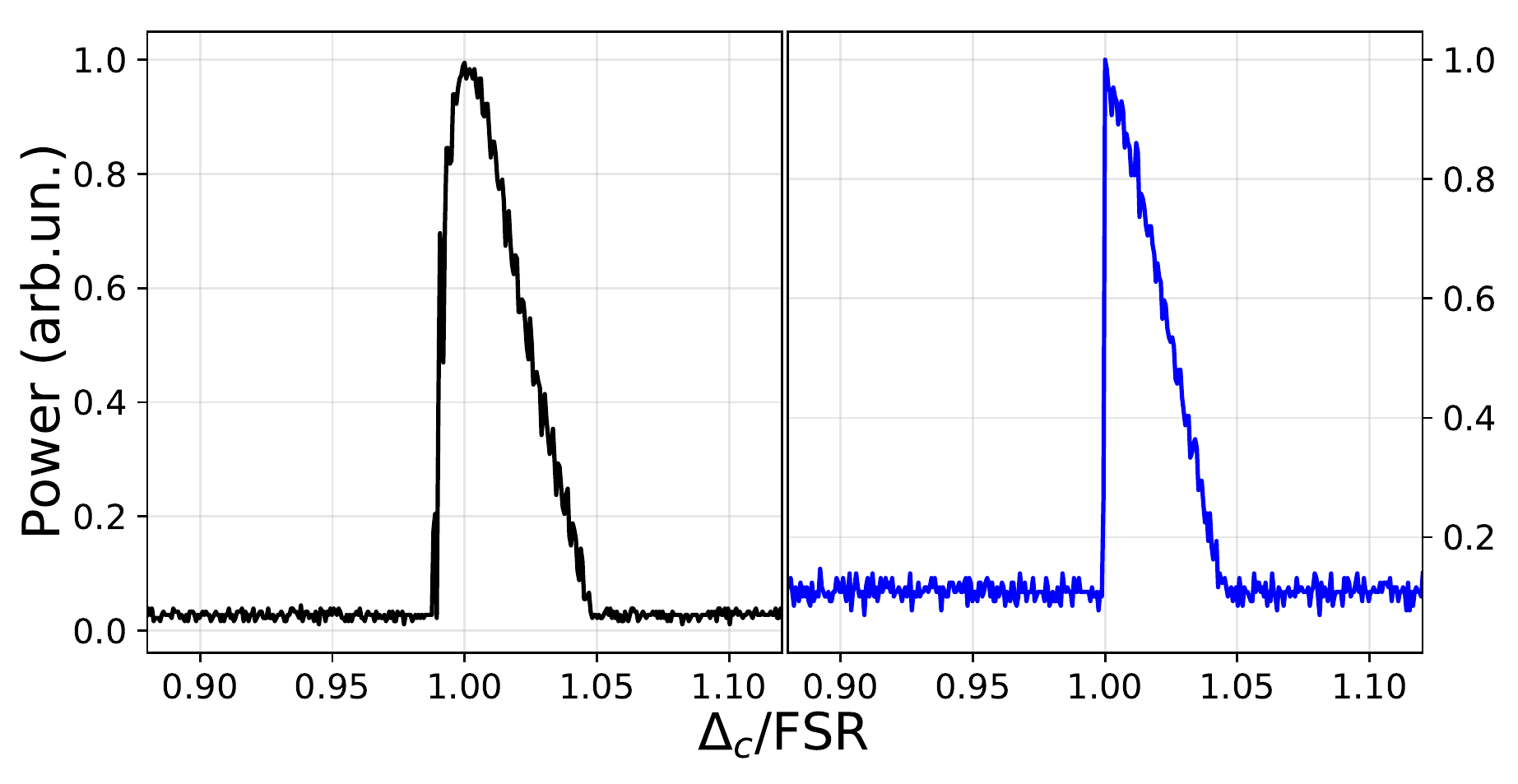}
\put(12,45){(a)}\put(54,45){(b)}
\end{overpic}
\caption{Output power from the OPO for a cavity length scan, with T=91$^{\circ}$C, $ \Delta= 1$
GHz, $\sigma=1.8$, for a finesse of (a) $\mathcal{F}=14$ and (b)  $\mathcal{F}=30$.}
\label{fig:OPO_output}
\end{figure}

Once the cavity is aligned, oscillation can be observed for sufficient pump power. During the scanning of the cavity length, whenever a doubly resonant condition for signal and idler beam is achieved, there is an intense output on PBS$_3$ (Figure \ref{fig:OPO_output}).
The sudden transition of the intensity while scanning the cavity length shows an abrupt threshold for the oscillation.
Figures \ref{fig:OPO_output} (a) and (b) show the recorded  output for two different cavity finesse, $\mathcal{F}=14$ and $\mathcal{F}=30$,  respectively, at the same normalized pump power $\sigma=P/P_{th}$ (where $P_{th}$ is the oscillation threshold pump power at exact resonance).

The asymmetry on the peak is a consequence of  the self-phase modulation associated to the  $\chi^{(3)}$ nonlinearity.
Close to the cavity resonance, the intensity  leads to a change of the field phase, thus either increasing or decreasing the absolute value of the  cavity detuning.  
For sufficiently high finesse (or for enough intracavity power) a bistable operation is observed, as a consequence of the atom-cavity cooperativity. For a two-level system this cooperativity is given by $C=g^2N/\gamma_{cav}\Gamma$
with $g$ as the atom-light coupling coefficient, $N$ the number of atoms, $\gamma_{cav}$ cavity loss rate and $\Gamma$ the atomic spontaneous emission ~\cite{Lambrecht95}.
 Notice that the increase of finesse, which reduces $\gamma_{cav}$,  enhances the atom-cavity cooperativity, leading to a the steep bi-stability of (b) with respect to figure (a). We work out of the bi-stability regime by  carefully choosing the value of $\mathcal{F}$.

\begin{figure}[h!]
\centering
\begin{overpic}[width=86mm]{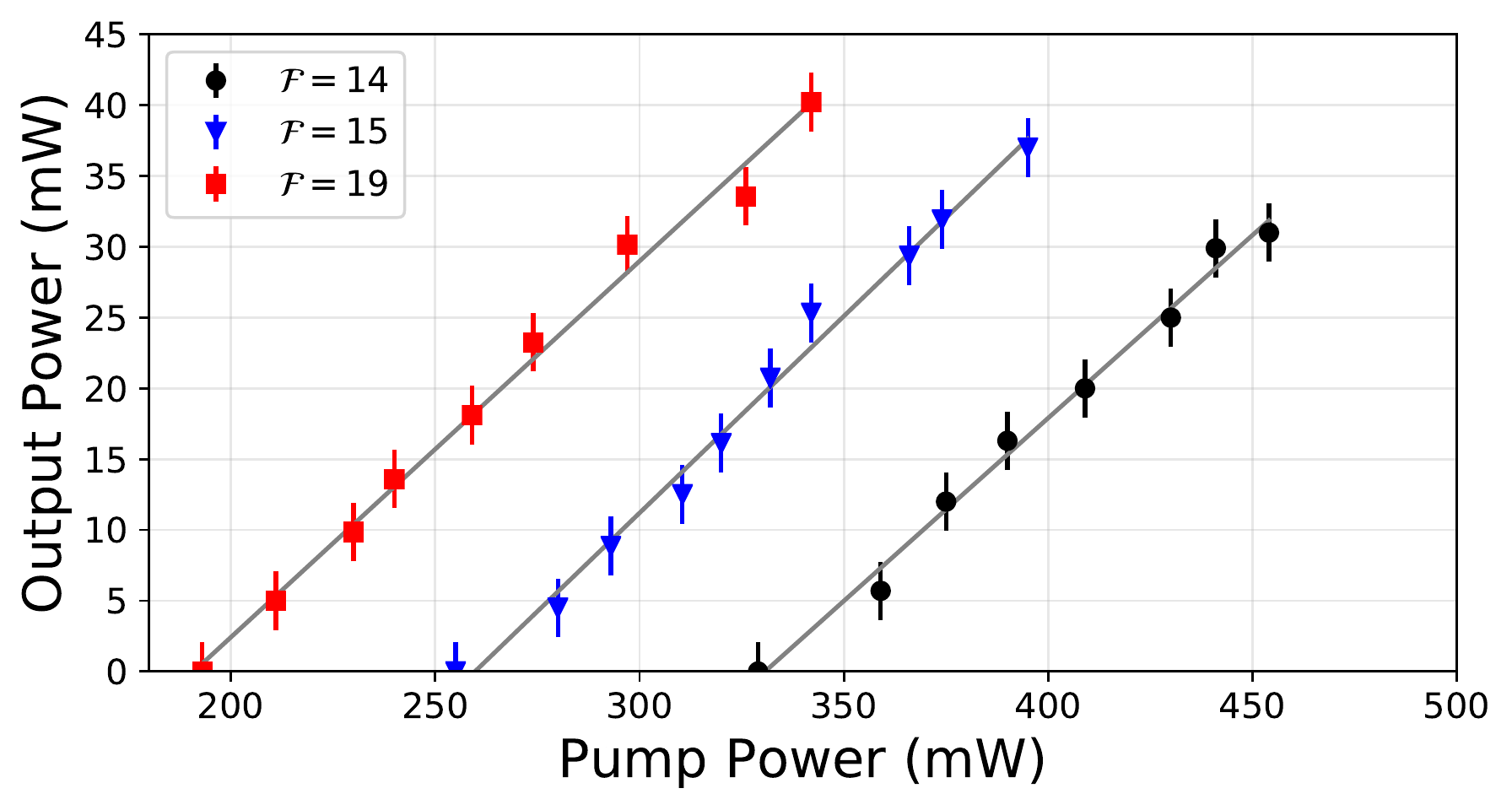}
\end{overpic}
\caption{Total output power as a function of the input pump power for three different values of finesse $\mathcal{F}$. T=91$^{\circ}$C, $\Delta= 1$GHz. The linear fit is just a guide to the eye.}
\label{fig:OPO_threshold}
\end{figure}

The high gain of the medium allows the oscillation with relatively large intracavity losses (of the order of 30 $\%$) and a controlable tuning of the optical coupling, which allows us to have a fine control of the OPO threshold. Figure  \ref{fig:OPO_threshold} shows the total output power as a function of the input pump power for three different values of finesse.  One can observe that the threshold power increases from 193 mW to 329 mW as the finesse of the cavity is reduced from $\mathcal{F}=19$ to $\mathcal{F}=14$ . 
That is significantly smaller than the typical operational condition of OPOs using $\chi^{(2)}$ media \cite{Villar05}. This condition could be reached due to the high gain of the medium, typically from 100\% to 500\% gain, going up to 2000\% in some cases \cite{Glorieux10}. In the oscillation regime, we should reach the saturation of the gain, matching the cavity losses \cite{dropoteo}. The high gain allows the operation with extremely lossy cavities.
Avoiding the strong bistable condition, we could lock the cavity to the resonance peak, obtaining stable operation of the OPO output with power ranges from 1 mW to 40mW, maintaining a single spatial mode. This is at least 10 times higher that a free-space 4WM parametric amplification, which typically runs at $\leq 1\ $ mW \cite{Marino08}. The output power can be increased up to  100 mW or more by increasing the pump power, nevertheless, exciting higher transverse electromagnetic modes, which its description is not  within the scope of this letter.

\begin{figure}[h!]
    \centering
    \includegraphics[width=86mm]{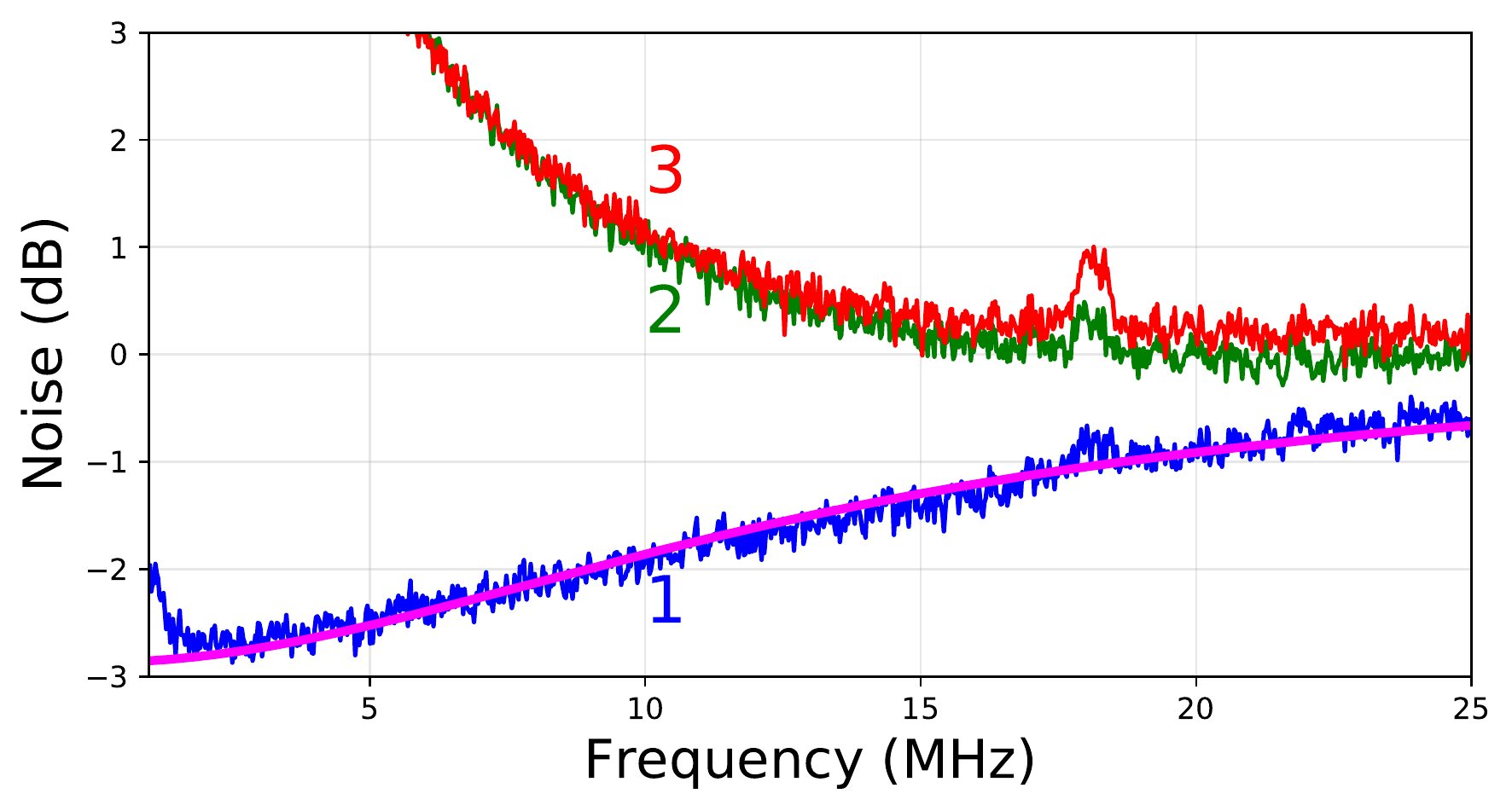}
    \caption{Intensity noise spectra of the output beams of the OPO, normalized to the shot noise level, for the subtraction of the photocurrents (1), signal (2) and idler (3) beams. T=91$^{\circ}$C, $ \Delta= 0.82$
GHz, $\mathcal{F}=15$, $\sigma=1.8$, $P_{th}=159$ mW.} 
    \label{fig:OPO_squeezing}
\end{figure}

As we generate pairs of beams, with nearly equal average intensities,
we look for the twin beam generation by looking to the noise spectrum of the intensity difference. 
Figure \ref{fig:OPO_squeezing} shows the normalized intensity noise spectra for the output beams of the OPO, after subtraction of electronic noise. 
Curve 1 shows the noise of the intensity difference between the probe and the conjugate beams, 
showing a maximum two-mode intensity difference squeezing of -2.7 (1) dB with respect to the shot-noise level at 2 MHz, for a pair of output fields with 12 mW of power for each beam. 
Experimental data can be compared  the simplest model for twin photon production inside a leaky cavity \cite{Reynaud}. We fit the noise spectrum by the Lorentzian response
\begin{equation}
    S(f)=1-\eta\frac{1}{1+(f/BW)^2}
\label{sqz}
\end{equation}
where $BW$ stands for the cavity bandwidth, and $\eta$ is the efficiency of the escape ratio of the photon through the output coupler $\eta=L_c/(L_c+L_i)$, where $L_c$ is the output coupler loss and $L_i$
corresponds to intrinsic cavity losses (adding up to 17\% in the present case). The resulting curve gives a maximum noise compression of -2.84 (16) dB, with a bandwidth of 16.1 (1) MHz, significantly smaller than the cavity bandwidth of 26 MHz.

The match of this simple model to the current result is quite surprising if we consider that the noise spectra of the twin beams generated from parametric amplification on atomic vapour can have a rich spectra \cite{Corzo13}, with twin beam correlation  shifting from squeezing to excess noise in the frequency range that is of the order of the atomic spectral linewidth of 6 MHz for the atom \cite{PDLett07}, depending on power broadening as well. The resulting profile is a consequence of the interplay of an amplifier with a variable gain bandwidth with a fixed cavity bandwidth, that cannot be fully accounted by simplified models where a broadband amplifier is considered \cite{dropoteo}.


Curves 2 and 3 in Figure \ref{fig:OPO_squeezing} show the normalized intensity noise for each beam. As it is expected, the noise for each beam presents excess of noise, and should be subjected to a more detailed treatment. 
Nevertheless, some distinctive features appear in this situation. The noise rapidly diverges for small analysis frequency, consistent with the cavity dynamics of an OPO, as a consequence of the phase diffusion between the converted fields  \cite{dropoteo}. This feature is absent in the case of the injected OPA \cite{PDLett07}, where the excess noise is not so sharp. That is also a consequence of the possibility of perfect squeezing of twin beams for lossless cavities, leading to strong excess noise in each field, differently from the finite squeezing level that is expected from single pass in a parametric amplifier.
Oddly enough, the noise of those fields goes down as the analysis frequency grows. In fact, it is expected that it could eventually evolve to noise compression for appropriate pump power under a certain frequency range \cite{dropoteo}.

\begin{figure}[h!]
    \centering
    \begin{overpic}[width=86mm]{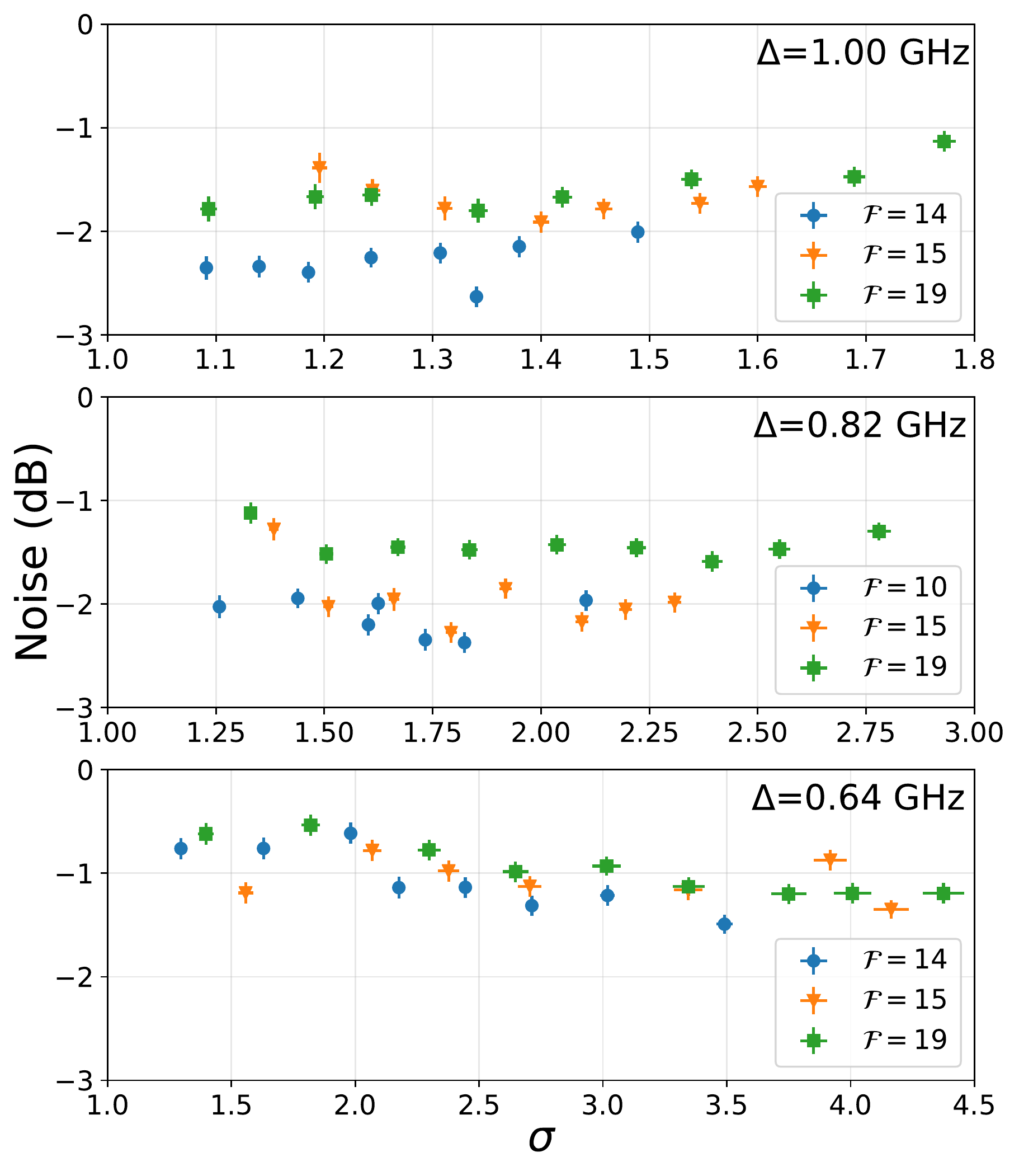}
    \put(10.5,95){(a)}\put(10.5,63){(b)}\put(10.5,11){(c)}
    \end{overpic}
    \caption{Noise intensity difference as a function of the input pump power normalised by the threshold power $\sigma$, for three different cavity finesse and an analysis frequency of  $7$ MHz.  (a), (b) and (c) correspond to three different pump detuning relative to the atomic line.} 
    \label{fig:OPO_squeezing_Ch_sigma}
\end{figure}

The present configuration provided a versatile tool to study the squeezing generation for different cavity couplings, adjusted by the intracavity waveplate. It changes both the loss ratio of the photon pair and the cavity bandwidth. Amplifying gain could be changed as well, by the control of the pump detuning.
Figure \ref{fig:OPO_squeezing_Ch_sigma}(a-c) shows the squeezing level for pump detuning frequencies of $\Delta=1.00,\, 0.82$ and 0.64 GHz.
We perform this characterisation for three different finesse values and as a function of the pump power above threshold. 
We can notice that, consistent with the expected results from either a simplified \cite{Reynaud} or detailed model \cite{dropoteo}, the squeezing level of the twin beams is insensitive to the pump power. This demonstrates also the robustness of the noise compression under the depletion of the pump field, even in the doubly resonant condition.
Other expected behavior is the reduction of the squeezing if we reduce the coupling of the cavity, that is consistent with the simple model shown in eq. \ref{sqz}.

A curious feature comes from the role of the amplifier gain. When the pump field is tuned closer to the atomic resonance, at 0.64GHz, the squeezing level is reduced to -1dB, due to the increase of the absorption in the atomic medium. On the other hand, since we have a higher gain, we could have lower threshold powers, and we demonstrate squeezing in values of $\sigma$  as high as 4.5 times above the threshold. As for the squeezing level, the best result is obtained for a lower cavity gain, at 1 GHz, for an open cavity, ranging from -2 to -3 dB. We expect that for reduced cavity losses the present configuration could provide a squeezing level at least as good as those obtained from the direct parametric amplification \cite{Glorieux10}.

\begin{figure}[h!]
    \centering
    \includegraphics[width=86mm]{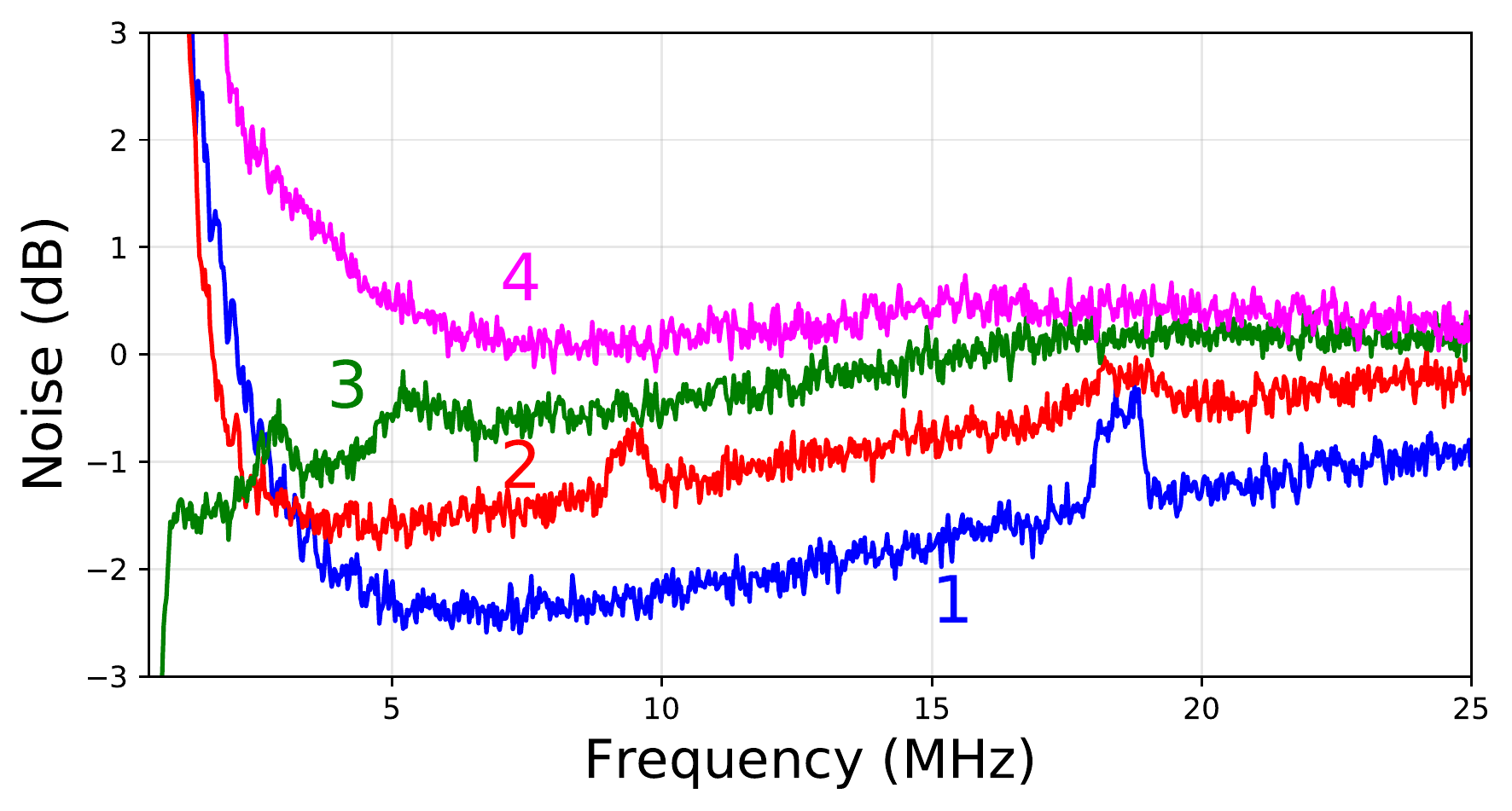}
    \caption{Noise spectra for different temperatures of the vapor cell. T=91$^{\circ}$C (Line 1),96$^{\circ}$C (L2),101$^{\circ}$C (L3),108$^{\circ}$C (L4) with $\sigma=1.8$, $\Delta=0.82$ GHz}
    \label{fig:OPO_temp}
\end{figure}

While finesse and detuning with respect to the atomic transition seem to keep the quantumness 
of the outcoming fields, within the range of parameters, the atomic density is a determinant parameter for twin beam generation.
Figure~\ref{fig:OPO_temp} shows the OPO noise spectra for different temperatures of the vapour cell, for the same normalized pump power $\sigma$. Notice that curve 1 for $91^{\circ}$C shows the maximum level of squeezing, and as the temperature is increased up to $108^{\circ}$C, the quantum correlations are lost and eventually the intensity noise difference is above the shot-noise. 
In other words, the increase of atomic density will increase the gain but should lead to an increase of the losses as well,
which deteriorate the quantum correlations of the twin beams. This could be the parameter that inhibited the observation of quantum correlations in previous realisations ref.  \cite{Okuma09,Sheng12}, since the temperature typically used is around  105$^{\circ}$C, which correspond to a noise spectrum in between curves 3 and 4 in Fig.~\ref{fig:OPO_temp}.

An interesting effect shown in Figs. \ref{fig:OPO_squeezing} and \ref{fig:OPO_temp} is a small peak, that appears in the noise of each individual beam, and is not suppressed by the subtraction. We performed a characterization of its frequency, and we have shown \cite{suppl} that it is proportional to the pump power. Although we lack a proper model for its origin, this dependence leads to an association with the AC Stark shift. Since it is narrow enough its presence doesn't affect the overall noise profile of the OPO output.

The OPO based on 4WM with hot atomic vapour has shown a significant level of squeezing in the twin beams (-2.7 (1) dB, -3.7 (1) dB after correction for quantum efficiency), that could be immediately optimized with the design of a dedicated cavity. This design is simplified by the possibility of using a relatively high transmittance for the output coupler ($\simeq 30\%$) while dramatically reducing the intracavity loss.
Moreover, since we are free from thermal effects and defective absorption of typical nonlinear crystals, an extended operational range of pump power could be studied, reaching more that four times above threshold before transverse multimode operation shows up. That allows the production of quantum correlation of very intense fields, with more than 20 mW each, a situation very distinct from the usual OPA amplifiers using this medium. Measurements above this power were limited by saturation on the photodetectors, so we don't discard that squeezing should remain for even higher intensities.
In the case of transverse multimode operation, an aperture in the smaller waist can be used to either select or manipulate those modes, leading to the production of fields featuring interesting quantum images. 

The presented results agree with the model of a gain medium in an open cavity, beyond the limit of week coupling \cite{dropoteo}. In this case, the pump depletion could be fully considered, and it was shown that, while it does not affect the robust twin beam generation, it leads to interesting dynamics of the noise compression of each beam, including here the depleted pump.
 Some interesting features, as the precise interplay between the bandwidths of the atomic amplifier and the cavity, or the origin of the narrow peak whose frequency closely follows the proportionality to the pump remain the subject of future studies.
We may conclude that the success in the observation of quantum correlations in the present implementation came from the control of the atomic density, operating at lower temperatures, and the use of a ring cavity, in order to avoid the effect or multiple coupling of propagating and counter-propagating modes in the atomic media. This system should provide a useful tool for the production of quantum correlated states close to atomic resonances, eventually leading to the observation of entanglement in such fields.


\section{Acknowledgements}
This work was funded by Grant  No. 2015/18834-0, 2017/27216-4 and 2018/03155-9  S\~ao  Paulo  Research  Foundation (FAPESP), and Grant No. N629091612184 (NICOP).

%
\end{document}


\preprint{APS/123-QED}

\title{Quantum noise correlations of an optical parametric oscillator based on a non-degenerate four wave mixing process in hot alkali atoms}
\author{A. Monta\~na Guerrero$^1$}
\author{P. Nussenzveig $^1$}
\author{M. Martinelli $^1$}
\author{A. M. Marino $^2$}
\author{H. M. Florez $^1$}
 \email{hans@if.usp.br}
\affiliation{$^1$ Instituto de F\'{\i}sica, Universidade de S\~ao Paulo, 05315-970 S\~ao Paulo, SP-Brazil}
\affiliation{$^2$ Center for Quantum Research and Technology and Homer L. Dodge Department of Physics and Astronomy, The University of Oklahoma, Norman, Oklahoma 73019, USA}

\date{\today}

\pacs{Valid PACS appear here}
\maketitle

\section{Data Analysis}
\begin{figure}[htb]
    \centering
    \includegraphics[width=86mm]{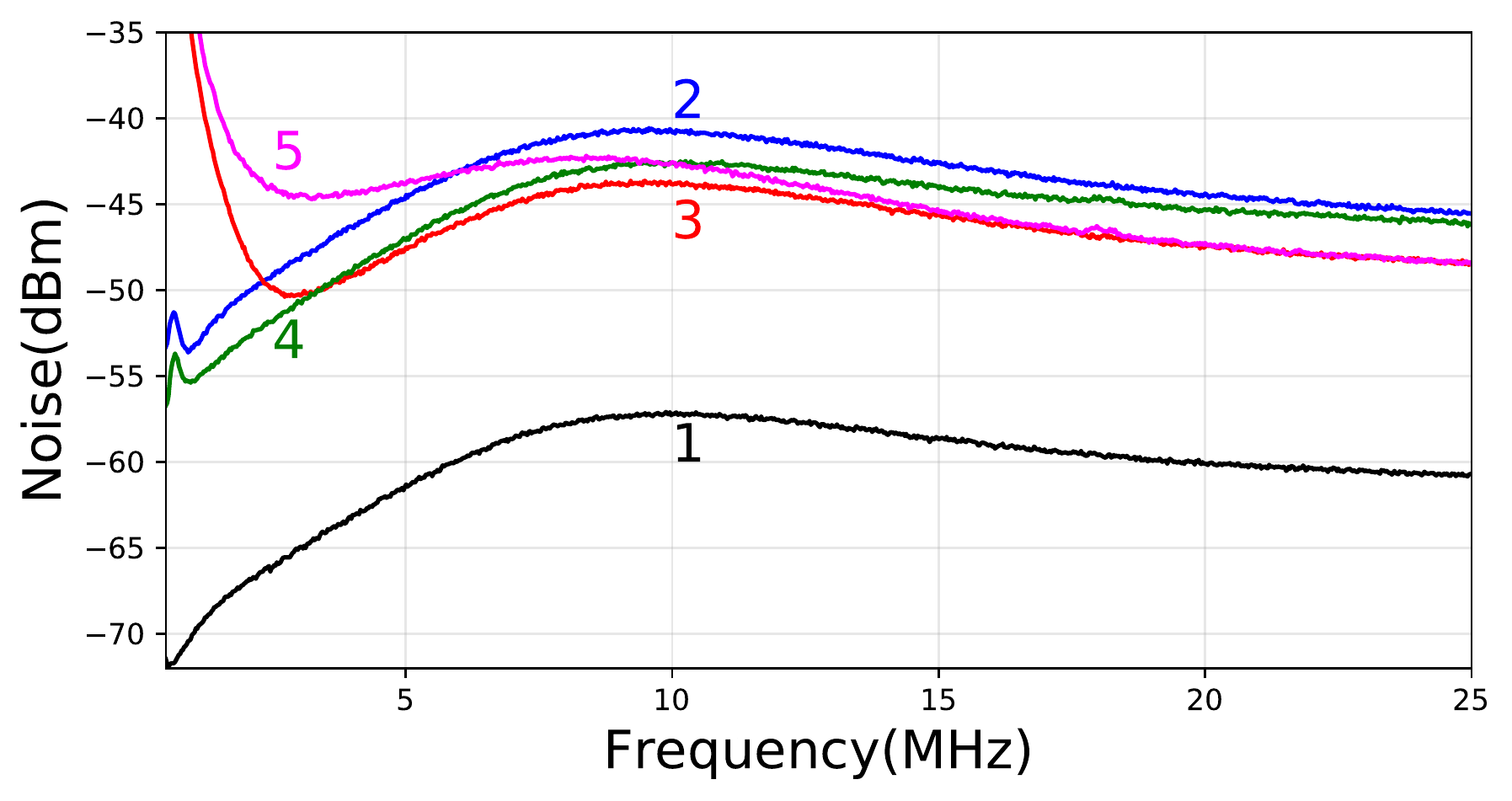}
    \caption{Measured noise spectra  used to evaluate the normalized curves on the main text. T=91$^{\circ}$C, $ \Delta= 0.82$
GHz, $\mathcal{F}=15$, $\sigma=1.8$, $P_{th}=159$ mW. Spectrum analyser settings: resolution bandwidth = 100 kHz, video bandwidth = 1 kHz.}
    \label{fig:datoBruto}
\end{figure}

Fig. \ref{fig:datoBruto} shows an example of raw data in which line 1 corresponds to the electronic noise, line 2 to the shot noise associated with the sum of the power of the two beams, line 3 to the shot noise associated with a single beam, line 4 to the noise of the intensity difference between signal and idler beams without normalization, and line 5 to the intensity noise associated to a single beam (signal or idler).
 
For the evaluation of the normalized noise presented in the main paper, all the measured spectra had their electronic noise subtracted before normalization by their corresponding shot noise level. This reference level was obtained using a sample of the pump laser, strongly attenuated for obtaining the same intensity of the output fields generated by the OPO. We have tested the intensity noise of the laser, verifying the linear response of the noise spectra with the power of the reference beam within a precision of 1\%.

\begin{figure}[h!]
    \centering
    \includegraphics[width=86mm]{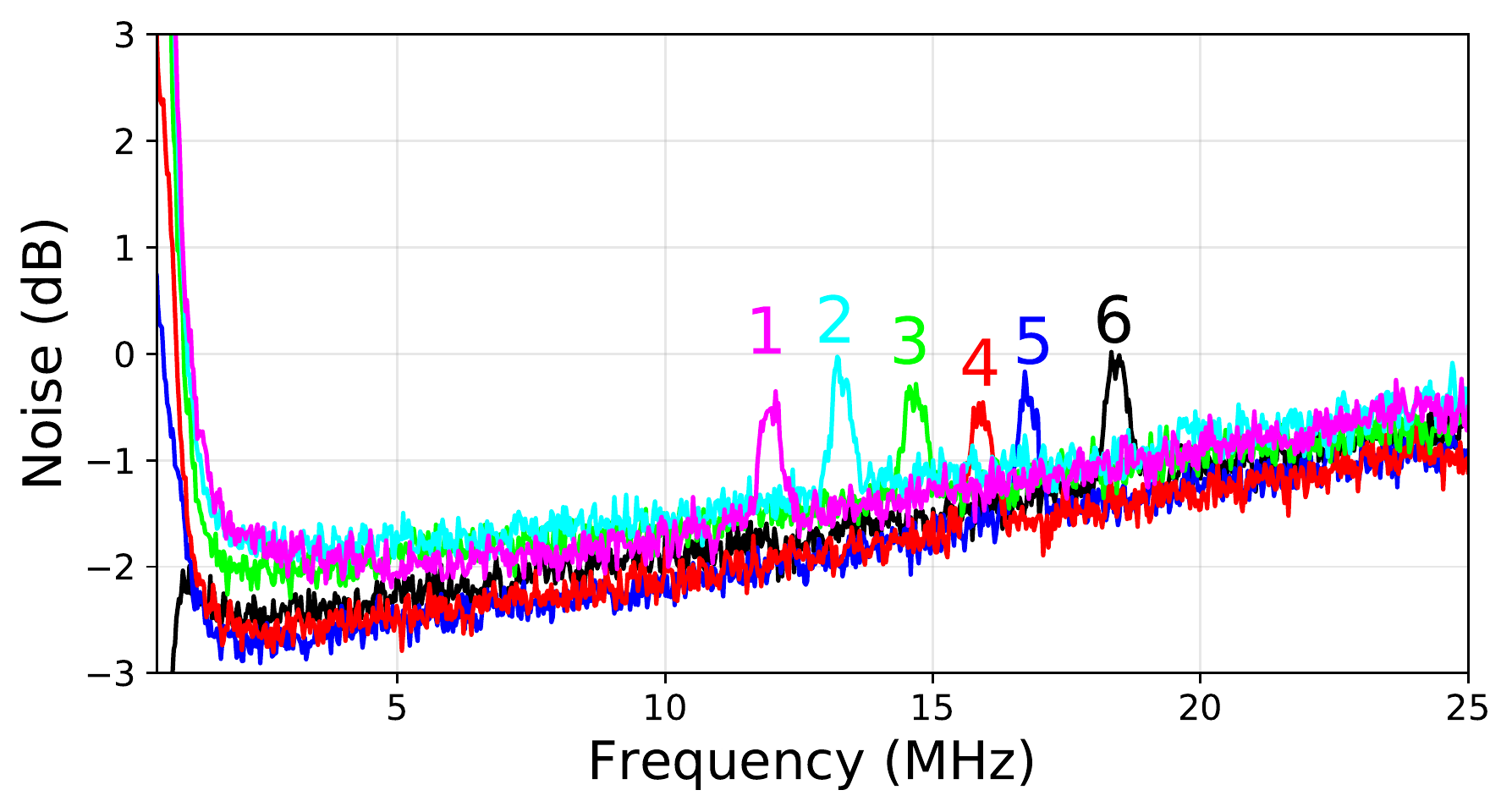}
    \caption{Normalized noise of the subtraction of the photocurrents. Curves 1 to 6 correspond to a input pump power of $\sigma$= 1.26, 1.44, 1.62, 1.73, 1.82 and 2.1mW respectively, all of them with $P_{Th}=$221mW. $\mathcal{F}=10$, $\Delta=0.82$ GHz,T=92$^{\circ}$C}
    \label{fig:RuidoVsFrecuenciaT}
\end{figure}

The evolution of the narrow peak in the noise spectra for different pump powers can be seen in Fig. \ref{fig:RuidoVsFrecuenciaT}. We present the normalized power spectrum of the intensity difference between signal and idler. Other parameters are the same as in Fig. \ref{fig:datoBruto}. Curves 1 to 6 correspond to an input pump power of 278, 318, 359, 383, 403 and 465 mW respectively.

\begin{figure}[h!]
    \centering
    \includegraphics[width=86mm]{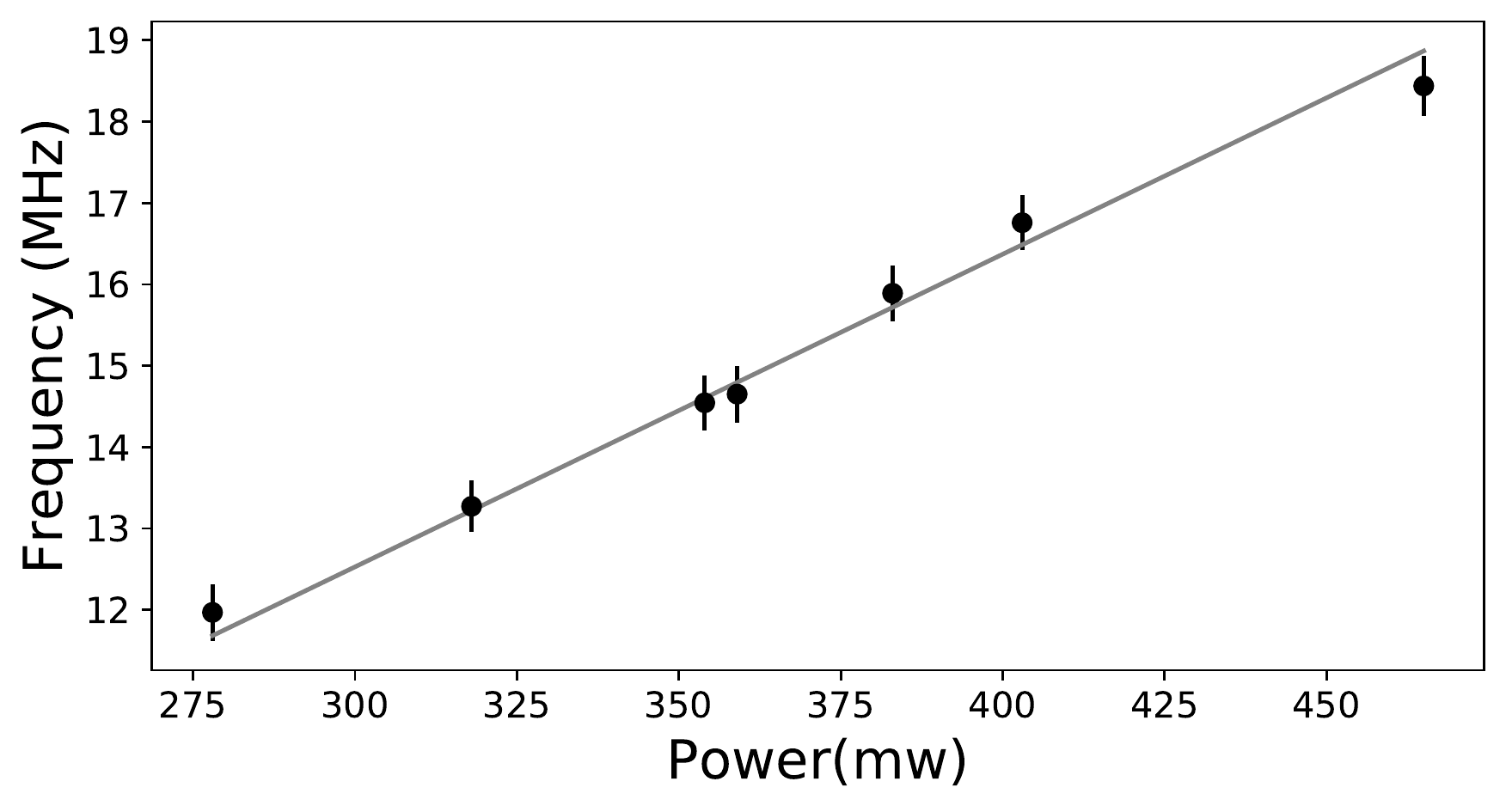}
    \caption{Peak frequency as a function of the pump power}
    \label{fig:my_label2}
\end{figure}

The peak frequency was evaluated by subtraction of the background noise taken as a baseline, followed by a fit to a Lorenztian. Central frequency is shown in Fig. \ref{fig:my_label2}, as a function of the pump power, while the error bars are estimated from the width of the fitted curve. 
The linear adjust corresponds to proportionality ratio of 38.4 (3) MHz/W.

